\input psfig.sty
\documentstyle[graphics,color,twocolumn,mnras_cite]{mn}
\def\spose#1{\hbox to 0pt{#1\hss}} 
\def\simlt{\mathrel{\spose{\lower 3pt\hbox{$\mathchar"218$}}
     \raise 2.0pt\hbox{$\mathchar"13C$}}}
\def\simgt{\mathrel{\spose{\lower 3pt\hbox{$\mathchar"218$}}
     \raise 2.0pt\hbox{$\mathchar"13E$}}}

\newcommand{\til}{$\sim$}

\newcommand{\msun}{\thinspace\hbox{$M_{\odot}$}}

\newcommand{\gpcmsq}{\nobreak\ifmmode{\thinspace\hbox{g}\thinspace\hbox{cm}^{-2}}%
	\else{$\hbox{g}\thinspace\hbox{cm}^{-2}$}\fi}

\newcommand{\beqn}{\begin{equation}}
\newcommand{\eeqn}{\end{equation}}
\def\spose#1{\hbox to 0pt{#1\hss}}
\def\simlt{\mathrel{\spose{\lower 3pt\hbox{$\mathchar"218$}}
     \raise 2.0pt\hbox{$\mathchar"13C$}}}
\def\simgt{\mathrel{\spose{\lower 3pt\hbox{$\mathchar"218$}}
     \raise 2.0pt\hbox{$\mathchar"13E$}}}
\def\degree{\nobreak\ifmmode{^\circ}\else{$^\circ$}\fi}
\def\arcmin{\nobreak\ifmmode{'}\else{$'$}\fi}
\def\arcsec{\nobreak\ifmmode{''}\else{$''$}\fi}
\def\hour{\nobreak\ifmmode{^{\rm h}}\else{$^{\rm h}$}\fi}
\def\minute{\nobreak\ifmmode{^{\rm m}}\else{$^{\rm m}$}\fi}
\def\second{\nobreak\ifmmode{^{\rm s}}\else{$^{\rm s}$}\fi}
\def\degreedot{\nobreak\ifmmode{^\circ\hskip-0.40em.\hskip0.08em}%
                         \else{$^\circ\hskip-0.40em.\hskip0.08em$}\fi}
\def\arcmindot{\nobreak\ifmmode{'\hskip-0.30em.\hskip0.02em}%
                         \else{$'\hskip-0.30em.\hskip0.02em$}\fi}
\def\arcsecdot{\nobreak\ifmmode{''\hskip-0.45em.\hskip0.08em}%
                         \else{$''\hskip-0.45em.\hskip0.08em$}\fi}
\def\seconddot{\nobreak\ifmmode{^{\rm s}\hskip-0.35em.\hskip0.05em}%
                         \else{$^{\rm s}\hskip-0.35em.\hskip0.05em$}\fi}
\def\flux{\nobreak\ifmmode{\times10^{-16}\,{\rm erg}\,{\rm cm}^{-2}\,{\rm s}^{-1}}%
	\else{$\times10^{-16}\,{\rm erg}\,{\rm cm}^{-2}\,{\rm s}^{-1}$}\fi}
\def\fluxw{\nobreak\ifmmode{\times10^{-19}\,{\rm W}\,{\rm m}^{-2}}%
	\else{$\times10^{-19}\,{\rm W}\,{\rm m}^{-2}$}\fi}

\def\eps@scaling{.95}
\def\epsscale#1{\gdef\eps@scaling{#1}}

\def\plotone#1{\centering \leavevmode
    \epsfxsize=\eps@scaling\columnwidth \epsfbox{#1}}

\title[Evidence for a 2 hr Modulation in GS1826-24]
{Evidence for a 2 hr Optical Modulation in GS1826-24}
\author[L. Homer et al.]
	{L. Homer$^1$, P. A. Charles$^1$, D. O'Donoghue$^{2,3}$\\
$^1$Department of Astrophysics, Nuclear Physics Lab., Keble Road, Oxford OX1 3RH\\
$^2$Department of Astronomy, University of Cape Town, Rondebosch 7700, Cape Town,South Africa\\
$^3$South African Astronomical Observatory, PO Box 9, Observatory 7935, Cape Town, South Africa.}

\date{Accepted 1999 December 31. Received 1999 January; in original
form \today}

\begin{document}

\maketitle

\begin{abstract}
We report the discovery of a 2.1hr optical modulation in the transient source
GS1826-24, based on two independent high time-resolution photometric
observing runs.  There is additional irregular variability on shorter timescales.  The source also exhibited an optical burst during each
observation, with peak fluxes consistent with those of the three X-ray bursts
so far detected by {\it Beppo}SAX.  We compare the low-amplitude variation
($\sim 0.06^m$) to that seen on the orbital periods of the short period X-ray
bursters, X1636-536 and X1735-444, as well as the similarity in their
non-periodic fluctuations.  Other transient neutron star LMXBs possess short
periods in the range 3.8-7.1 hrs.  However, if confirmed as the orbital, a
2.1 hr modulation would make GS1826-24 unique and therefore of great interest
within the context of their formation and evolution.

\end{abstract}


\section{INTRODUCTION}
GS 1826-24 was discovered serendipitously in 1988 by the
{\it Ginga} LAC during a satellite manoeuvre (Makino {\it et al.} 1988). 
The source had an average flux level of 26 mCrab (1-40 keV), and a power law spectrum  with $\alpha = 1.7$. Observations both a month before and after by the {\it
Ginga} ASM, by TTM in 1989 \cite{intZ92} and by ROSAT in 1990 and 1992
\cite{barr95} found comparable flux levels.  Temporal analyses of both the {\it
Ginga} detection and ROSAT data yielded a featureless $f^{-1}$ power
spectrum extending from $10^{-4} - 500$ Hz \cite{tan95,barr95}, with
neither QPO nor pulses being detected.

Despite its detection by {\it Ginga}, the source had not been previously
catalogued.  Neither were X-ray bursts detected by {\it Ginga}.  Together
with its similarities to Cyg X-1 (hard X-ray spectrum, strong flickering),
this led to an early suggestion by \scite{tana89} that it was a soft X-ray
transient with a possible black-hole primary.  Later, \scite{stri96}
called this suggestion into doubt, following examination of data from
CGRO/OSSE observations. They found that fitting both the Ginga and OSSE
spectra produced a model with an exponentially cutoff power law plus
reflection term.  The observed cut-off energy of $\sim$58 keV is typical
of the cooler neutron star hard X-ray spectra.  The recent report of three X-ray bursts
detected by {\it Beppo}SAX \cite{uber97} and our detection of optical bursts here confirms the presence of a neutron star accretor. 

Following the first ROSAT/PSPC all-sky survey observations in September
1990, and the determination of a preliminary X-ray position, a search for
the counterpart yielded a time variable, UV-excess, emission line star
\cite{motc94,barr95}. This source had $B= 19.7 \pm 0.1$, and an uncertain
V magnitude of $V \simeq 19.3$, due to contamination by a nearby star.  There was also evidence for $\simeq 0.3^m$ variations on a one hour
timescale, but the time sampling was fairly poor. For this reason, 
we included this object in our target list for a high-speed photometry run at the South African Astronomical Observatory (SAAO). Further
time-series photometry
was also obtained on the William Herschel Telescope, La Palma (WHT) to confirm the variability that was seen.

\section{OBSERVATIONS AND DATA REDUCTION}
\subsection{SAAO 1996 June}
Observations of a small (50x33 arcsecs) region surrounding the optical counterpart to GS1826-24 were made using
the UCT-CCD fast photometer \cite{od95}, at the Cassegrain focus of the 1.9m telescope at SAAO, Sutherland on 1996 June 20.  The UCT-CCD fast
photometer is a Wright Camera 576x420 coated GEC CCD which was used here half-masked so as to operate in frame transfer mode, allowing
exposures of as short as 1s with no dead-time.  We observed from UT 00:03:27 to UT 04:15:15, obtaining 1507 consecutive 10s exposures. The
conditions were generally good, being almost photometric, and the mean seeing was $\sim$1.5 arcsec.  However, problems with telescope focus drift did lead to a
50 min
interval of poorer quality data.  In order to obtain the best signal-to-noise for the highest time resolution  we used no filters. 

\subsection{WHT 1997 July}
Observations of  GS1826-24 were made on 1997 July 31 using the auxiliary port camera of the WHT. In order to provide the highest  possible time
resolution the TEK chip was windowed down to a small 33x33 arcsecs field-of-view, giving a dead-time of 10s.  A series of 335 consecutive 25s B band
exposures were taken from UT 21:54 to UT 01:10.  Conditions were photometric throughout the run, and the seeing was
generally quite good, varying between 1 and 1.7 arcsecs, apart from the first 15 min with $\sim$ 2 arcsec.

\subsection{Photometry}
The imperfect auto-guiding of the SAAO 1.9m led to the positions of stars moving both randomly and with a definite trend
 during the run
 (from examination of coordinates of a bright star on the frame).  To cope with the motion of the stars relative to each frame, an  {\small
 ARK} routine ({\small MAPCCD}) was used to map the positions of about 10 bright stars on each image, compared to a reference image, and
 calculate initial offsets for star centres.  A further improvement was then made by using these initial offsets to find the centroids of three bright
 stars, producing the final offsets.  Similarly, two bright reference stars were used to calculate the small offsets relative to a
 reference image for the WHT run in one step.  

The following was then undertaken for both data sets:
\begin{enumerate}
\item The centres of all stars (above a chosen threshold) were measured on the reference image
 using the {\small FIND} routine in {\small DAOPHOT II} \cite{stet87}. The star list was edited to remove stars affected by any CCD defects
as well as those stars both too close to the edge and very faint, creating a master star list.
 This included a suitable number of bright local standards as well as  stars of comparable brightness to the counterpart.  

\item Due to moderate
 crowding of the counterpart with a nearby but fainter neighbour and the variable seeing, point spread function (PSF) fitting was essential to
 obtain good photometry. We performed the photometry reductions with the {\small IRAF} implementation of {\small DAOPHOT II}.  An automated PSF
 fitting set-up was employed, whereby the same well-isolated and bright stars (see Fig. \ref{fig:images}) were selected as PSF stars for each run, and the PSF
 was modelled on the profiles of these stars for each image separately.  Then for each star of interest the PSF was scaled according to an
 initial aperture photometry estimate and fitted in order to measure its brightness.

\begin{figure*}
\resizebox{0.547\textwidth}{!}{\rotatebox{90}{\reflectbox{\includegraphics{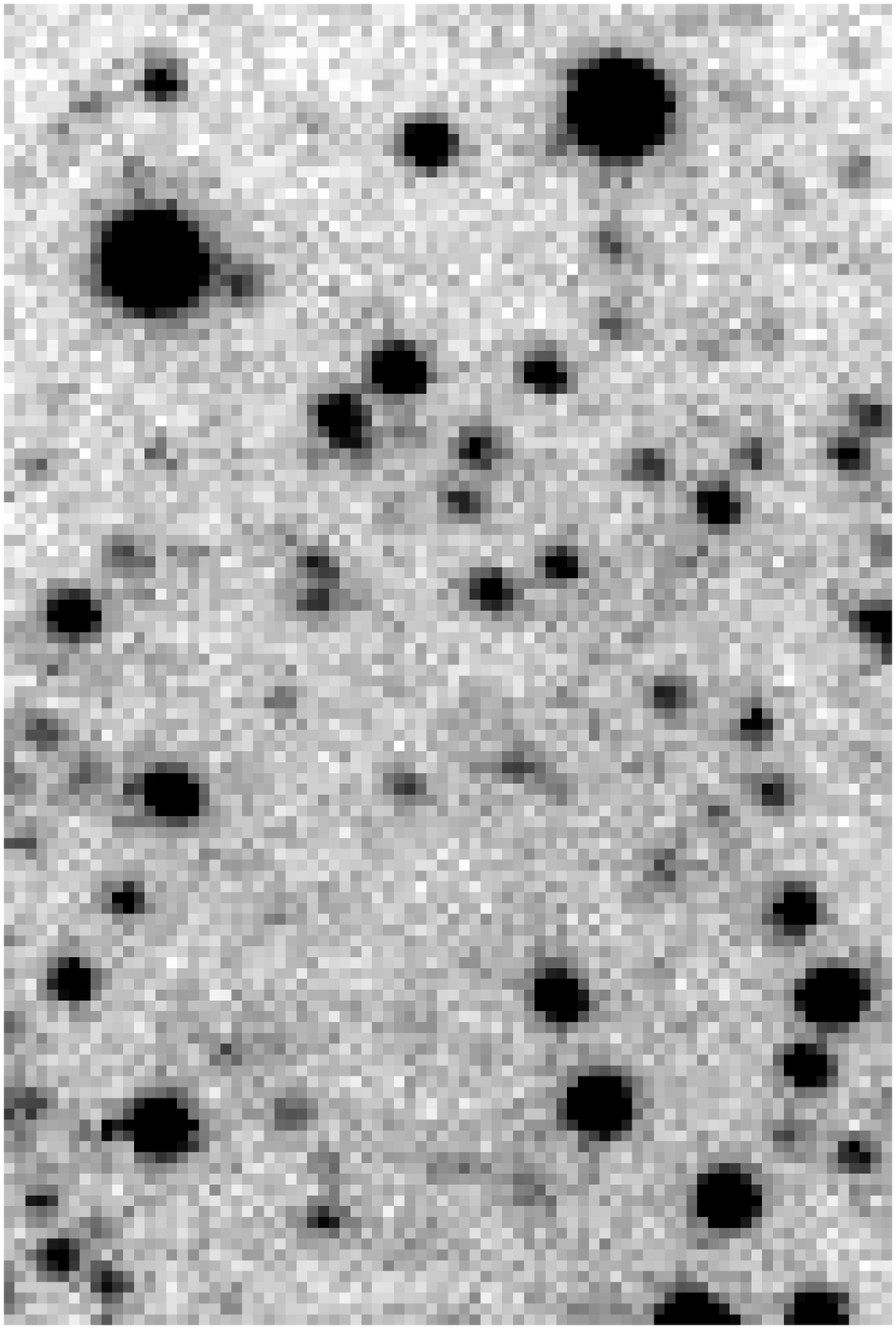}}}}
\resizebox{0.37\textwidth}{!}{\reflectbox{\includegraphics{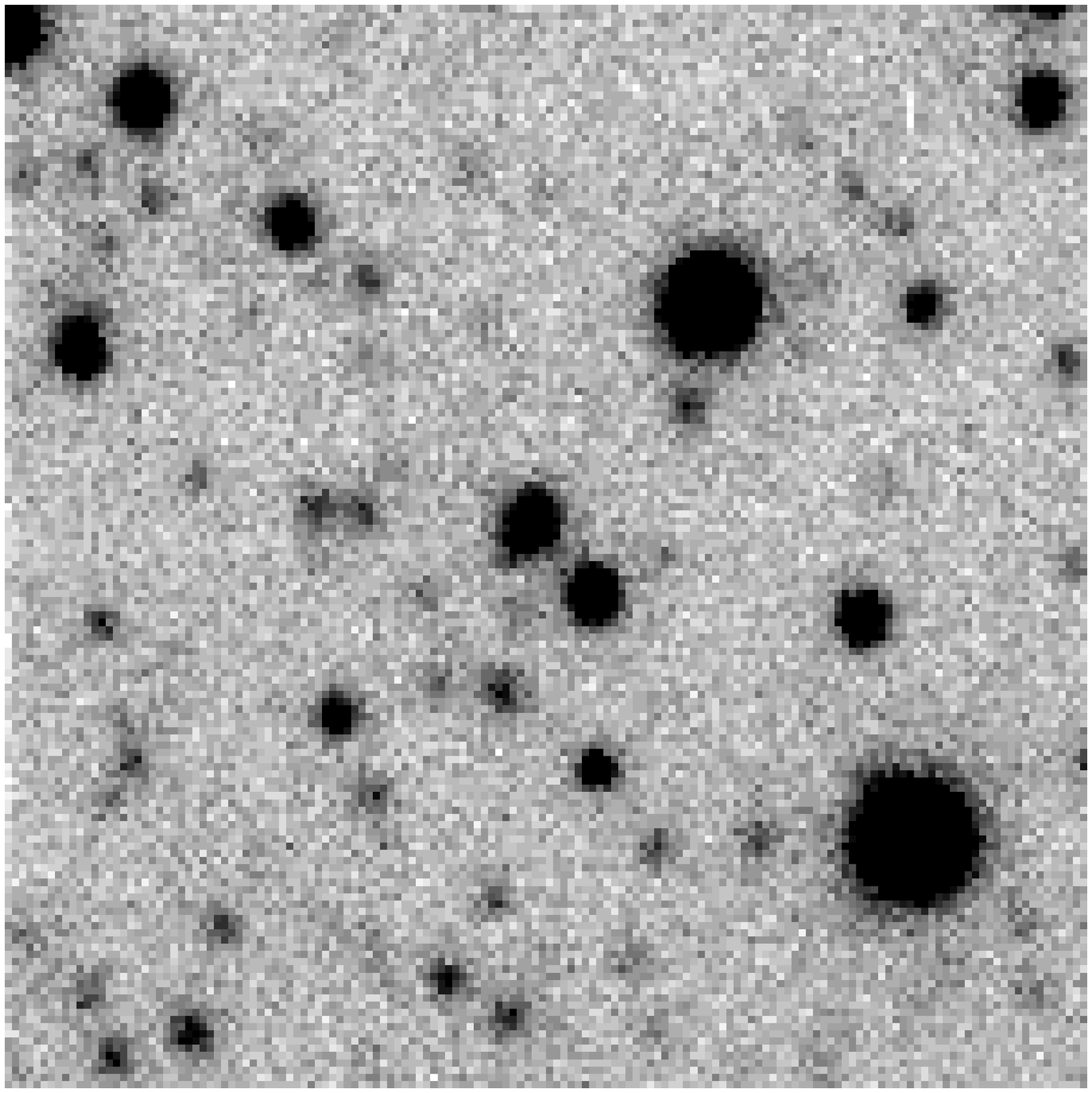}}}
\color{black}
\put(-400,167){\color{black} \small \textbf{PSF}}
\put(-433,70 ){\color{black}\small \textbf{PSF}}
\put(-225,144){\color{black}\small \textbf{PSF}}
\put(-230,56 ){\color{black}\small \textbf{PSF}}
\put(-119,147){\color{black}\small \textbf{PSF}}
\put(-149,60 ){\color{black}\small \textbf{PSF}}
\put( -88,103){\color{black} \small \textbf{OC}}
\put(-376,125){\color{black} \small  \textbf{OC}}
\put(-387,123){\textbf{$\backslash$}}
\put(-382,103){ \textbf{$\backslash$}}
\put(-103,105){ \textbf{$\backslash$}}
\put( -95,85 ){ \textbf{$\backslash$}}
\color{black}

\caption{Images showing the field of GS1826-24 (North up, East to the left): 10s white light exposure from SAAO June 1996 - 49x33 arcsecs field of view (left), 25s B band
 exposure from WHT 1997 - 33x33 arcsecs field of view (right).  The optical counterpart and the bright stars used for PSF fitting are labelled in each case.\label{fig:images}}
\end{figure*}

\item In order to double-check the psf fitting photometry, we also ran aperture photometry using a rectangular aperture, positioned to include both the counterpart
and its neighbour.  The resulting lightcurve was identical, apart from being somewhat noisier.

\item For this variability study only the relative brightness of a star is of importance, and so we applied differential photometry to help reduce
the effects of any variations in transparency, employing up to 6 bright local standards. 
\end{enumerate}

\begin{figure*}
\resizebox*{1.02\textwidth}{0.46\textheight}{\rotatebox{-90}{\includegraphics{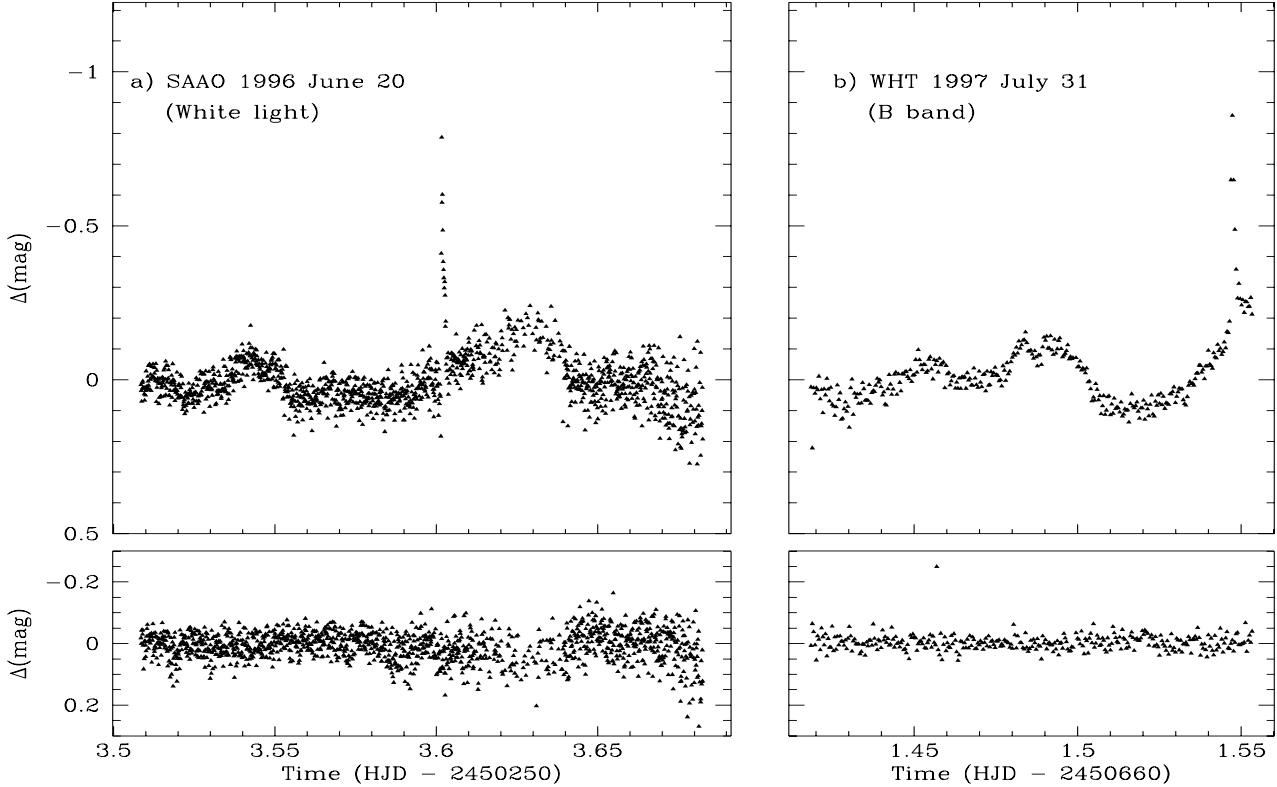}}}
\caption{Lightcurves of GS1826-24 and star of comparable brightness. a) High speed photometry in white light from SAAO with 10s resolution,
b) conventional B band photometry from WHT with 35s time resolution.\label{fig:1826lc}}
\end{figure*}

Owing to the variable seeing conditions and problems with telescope focus drift during the run, a small portion of the SAAO data is of
poor quality.  Hence, before subsequent analysis those data points corresponding to frames with estimated seeing $>$2.5 arcsecs were deleted.  The resulting  4.19 hour and 3.26 hour duration lightcurves for the optical counterpart to GS1826-24 and a suitable star of comparable brightness are shown in
Fig.\ref{fig:1826lc}, with a quadratic detrend applied.    Two features are immediately apparent; two very clear optical bursts, and evidence for short-term variations   in 
brightness.

\section{OPTICAL BURSTS}
\begin{figure*}
\resizebox{0.8\textwidth}{!}{\rotatebox{0}{\includegraphics{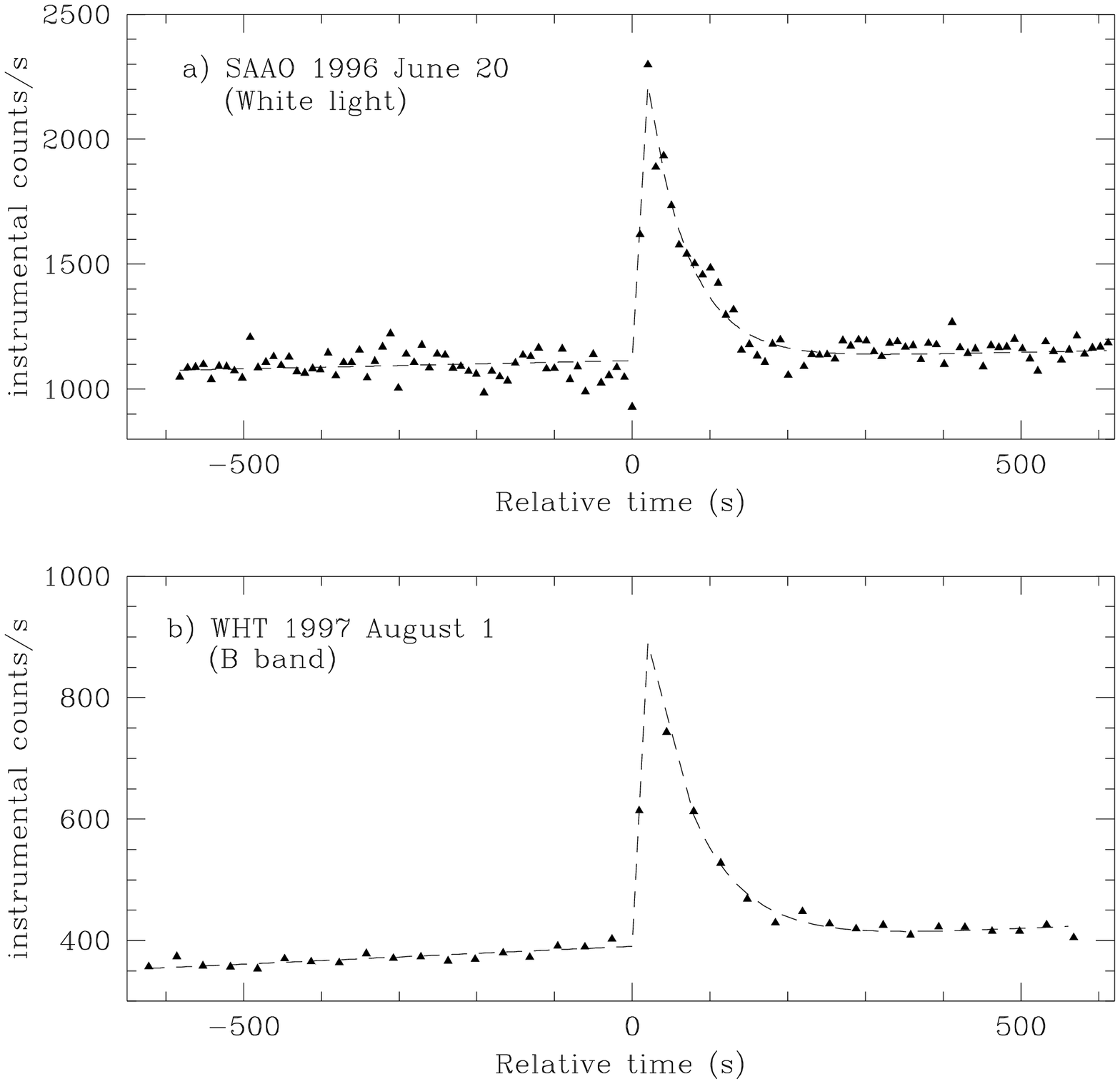}}}
\caption{Optical burst lightcurves (symbols) plus model fit (dashed line). a) High speed photometry in white light from SAAO with 10s resolution,
b) conventional B band photometry from WHT with 35s time resolution.  The limiting model parameters are adopted with a rise time of 20s in each
case.\label{fig:1826burs}}
\end{figure*}

A simple burst profile consisting of a linear rise, plus exponential decay was fitted to a 1200s section of each lightcurve including the bursts (as shown in
Fig.\ref{fig:1826burs}).  When fitting the burst a number of different models were used for the variable persistent flux;
a double sinusoid constrained to fit the entire lightcurve and  a linear  or a quadratic baseline over the 1200s section, however the burst parameters were
found to be insensitive to the model chosen.  The time sampling (10 and 25s at SAAO and WHT respectively) was adequate to constrain the decay of the burst, but
not the fast rise and hence peak flux.  We can therefore only place limits.  The fit to the SAAO June 1996 burst (reduced $\chi^2= 1.6 $ for 113 d.o.f.) yielded the 
following parameters: rise time $<20$s, e-folding decay time$=53.3\pm2.7$s, $\Delta F_{opt}=1100 {\rm cts\thinspace s}^{-1}$ (i.e. 2.0 x average flux).
Similarly for the WHT August 1997 burst  (reduced $\chi^2= 0.8$ for 29 d.o.f.), we obtained a rise time $<65$s, e-folding decay time$=69.5\pm4.3$s,
$\Delta F_{opt}=390{\rm cts\thinspace s}^{-1}$ (note: constraining the rise time to $<20$s gives $\Delta F_{opt}=500{\rm cts\thinspace s}^{-1}$, i.e. 2.4 x average flux).   The results are summarised in Table \ref{tab:1826burs}. 

\begin{table*}
\caption{Fitted parameters of a simple burst profile, with linear rise-time and exponential decay\label{tab:1826burs}}

\begin{tabular}{l l l l l l  } \hline
Date & Time (UT) & Rise time (s) & $\Delta F_{opt}({\rm cts\thinspace s}^{-1})$ & e-folding decay time (s) & reduced $\chi^2$ \\
\hline
20 June 1996 &  02:18 &  $<20$ & $1100$ & $53.3\pm2.7$ & 1.55  (113 d.o.f)
\\
1 August 1997 & 01:02  & $<65$ &  $390$ & $69.5\pm4.3$ &   0.8 (29 d.o.f.) \\
				&				& set $<20$ 		& $500$  & $69.1\pm2.8$ &   0.9 (29 d.o.f.) \\		
\end{tabular}

\vskip 5pt

\end{table*}

\begin{table*}
\begin{minipage}{150mm}
\vskip 5pt
\caption{Calculation of the peak X-ray burst flux corresponding to the two optical bursts of GS1826-24 \label{tab:burs-calc}}

\begin{tabular}{l l l l l l l } \hline
Date & Passband  & $\alpha$ & $\Delta F_{opt}({\rm cts\thinspace s}^{-1})$ & $\left(\frac{F_{opt,max}}{F_{opt,pers}}\right)$ & $F_{X,pers}$(mCrab)
\footnote{Estimates of persistent flux taken from the XTE/ASM 2-10keV one-day average on the observation date} & $F_{X,max}$(mCrab)\\
\hline
20 June 1996 & White light & 2.8   & $> 1100$ & 2.0 & 27 & $>190$ \\
1 August 1997 & B  &  2.55 &  $> 500$ & 2.35 & 31 &  $ >270$ \\
	
\end{tabular}

\vskip 5pt

\end{minipage}
\end{table*}

Many optical bursts have been detected from the X-ray burster X1636-536 (see e.g. \pcite{ped82}).  In general, they exhibit a wide range of peak
fluxes and profiles, with  $F_{max}/F_{pers} = 1.4 - 4.8$ ( in white light) and typical decay times \til few tens of
seconds.  Clearly, the bursts from GS1826-24 are not dissimilar.

Lastly,  we may compare the corresponding X-ray peak fluxes of these optical bursts to those of the bursts detected by  {\it Beppo}SAX
  \cite{uber97}, using the results of \pcite{law83}. They derived a simple power law relation  between the changes in the  U, B and V  band fluxes and corresponding X-ray flux variations
during  a well-studied
burst of X1636-536, with $\left(\frac{F_{X,max}}{F_{X,pers}}\right) = \left(\frac{F_{opt,max}}{F_{opt,pers}}\right)^\alpha$, where $\alpha$ varies with the passband.

  The results of these calculations are presented in Table \ref{tab:burs-calc}, where we use the $\alpha$ value for V as an approximation for
  the white light observations.  The peak fluxes measured by {\it Beppo}SAX range from 350 to 520 mCrab, hence our lower limits of 190 and 270
  mCrab imply consistency.




\section{SHORT PERIOD MODULATIONS}
 Firstly, we detrended the extracted light curves for each of the stars using quadratic fits. In the case of GS1826-24 the
data points corresponding to the burst were also excluded. The maximum of the subtracted polynomial was constrained to the time of least
airmass for the field, ensuring reasonable compensation for 
 the slow variations due to airmass changes.  To search for periodic modulations two different methods were employed:
(i) we calculated a Lomb-Scargle (LS) periodogram routine on each data set, to search for sinusoidal modulations (this periodogram is a modified
discrete Fourier transform (DFT), with normalisations which are explicitly constructed for the general case of time sampling, including uneven sampling, see \pcite{scar82});
(ii) we constructed a phase dispersion minimisation periodogram (PDM), which works well even for highly non-sinusoidal light curves (see \pcite{stell78}).

\begin{figure*}
\resizebox{1.0\textwidth}{!}{\rotatebox{-90}{\includegraphics{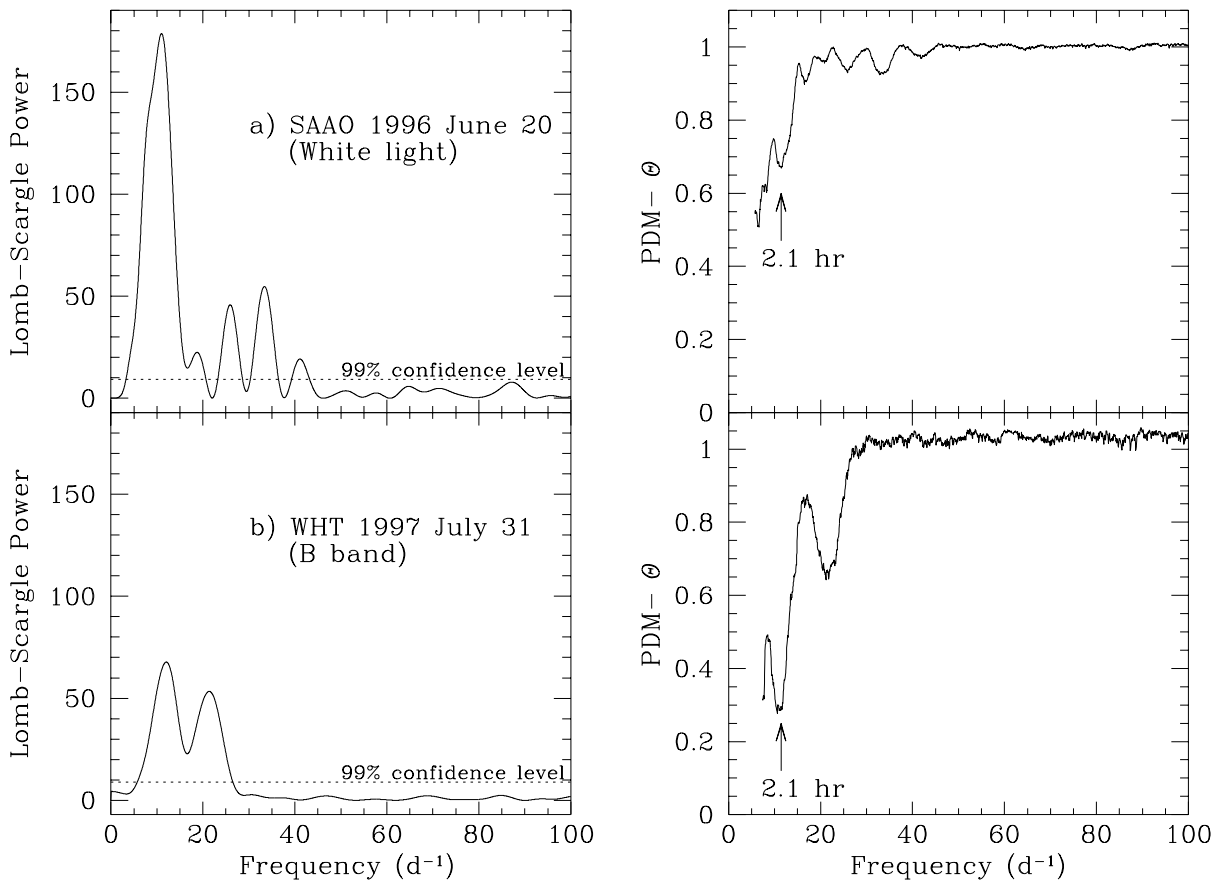}}}
\caption{Lomb-Scargle Periodograms (left) Phase-Dispersion Minimisation Periodograms (right) for GS1826-24 lightcurves from a) SAAO and b) WHT with
bursts excluded. The PDM plots have been truncated at the lowest frequencies corresponding to the reciprocal of the data duration.\label{fig:lsp-csm}}
\end{figure*}

The resultant LS periodograms for both GS1826-24 datasets show distinct peaks at close to $11.5$d$^{-1}$ ($P=2.1$ hr)
(see Fig. \ref{fig:lsp-csm}).  The exact measured frequencies are 11.0$\pm 0.2$d$^{-1}$ and 12.0$\pm 0.4$d$^{-1}$ for the SAAO and WHT
lightcurves respectively, with semi-amplitudes of 0.05$^m$ and 0.07$^m$ for a fitted sinusoid.  Although the frequencies are only marginally
consistent (at $\simeq 2 \sigma$ level)
within the formal random errors quoted, it must be noted that the datasets span only 2.0 and 1.6 cycles, with 
airmass changes of 1.0-2.4 and  1.6-2.0 respectively, and hence the precision of the period determination is limited to $\pm 0.8$d$^{-1}$
according to the exact form of detrending used. The formal
confidence of these peak frequencies are greater than 99.98 per cent, as determined from a cumulative probability
distribution (CDF) appropriate for the dataset (see \pcite{home96} section 3 for details of the method employed).  There are also
significant peaks at the higher frequencies of $25.9 \pm 2.3$d$^{-1}$, $33.4 \pm 0.25$d$^{-1}$ ($P=55.7 \pm 0.5$ min, $P=43.1 \pm 0.3$ min) [SAAO] and $21.4 \pm 0.45.$d$^{-1}$ ($P=67.3 \pm 1.6$ min)
[WHT] corresponding to the shorter timescale variations.  The results of PDMs confirm those of the Fourier analysis, with peaks occurring at
11.5d$^{-1}$ and 11.0d$^{-1}$ , corresponding to the longer periods
and also at 25.8d$^{-1}$, 32.8d$^{-1}$ and 21.2d$^{-1}$ (see Fig. \ref{fig:lsp-csm}).

We also examined the LSP results for stars of comparable brightness.
None of the four in the SAAO dataset exhibit any peak on 2-3 hr timescales with amplitude greater than 0.015$^m$, nor do these variations bear any phase
relationship with the 2hr modulation seen in GS1826-24.  Since the brightest stars in the
field also show such small scale variations, they are consistent with the small amplitude atmospheric transparency
variations found at SAAO, enhanced by the effect of poorer photometry during the interval of inferior seeing.

Similarly, the largest peak in the 2-3 hr range shown by the 3 check stars in the WHT dataset corresponds to a very low (0.007$^m$) amplitude
modulation.  Again this is most probably atmospheric in origin.
 
In any case, the presence of the 2hr modulation of GS1826-24 in {\it both} datasets provides convincing evidence of its reality.  Nevertheless, the short data span and the
complication of additional variability makes it impossible to constrain the stability of the modulation and hence identify it categorically as
orbital in origin.  Longer time base observations of this object are clearly necessary.

\section{DISCUSSION}
\begin{table*}
\begin{minipage}{150mm}

\caption{Properties of GS1826-24 and the short-period transient bursters$^a$\label{tab:sptb}}
\footnotetext[1]{Data taken from \scite{vP95} and references therein, unless cited elsewhere.}
\begin{tabular}{c c c l c l l l } \hline
Source 	& $P_orb$ & V 		& B-V, U-B 	&	E${\rm _{B-V}}$		& F$_X$	&	Active &Quiescent 			\\
			&	(hr)	&			&	& & ($\mu$Jy)&	&  \\
\hline
GS1826-24&	2\footnote[2]{This work.}		&19.3	\footnote[3]{\scite{motc94}.}	& 0.4$^c$, -0.5$^c$	&	0.4$^c$	& 30$^c$& 1988-present\footnote[4]{\scite{mak88}.} & before 1988\\
			&			&			&	&			&		&&\\
X0748-678&	3.82	&$>$23\footnote[5]{\scite{wad85}.}-16.9&0.1, -0.9	&0.42		& 0.1-60	& 1985-present\footnote[6]{\scite{whi95}.} &before 1985\\
			&			&			&		&		&	&	\\
X2129+470 &	5.24	&16.4-17.5&	0.65,-0.3 &0.5		&	9	& $<$1979 - 1983$^f$ & 1983-present\\
			&			&			&	&		&			&		&\\
X1658-298 & 7.11&  18.3		&0.45,-0.4		& 0.3 & $<$5-80& pre-1980s $^f$ & 1980-present	\\
			&			&	&	&			&		&&\\
\end{tabular}

\end{minipage}
\end{table*}


The period distribution of low-mass X-ray binaries (LMXB) has
shown a scarcity of systems below 3 hr presumably in part due to their faintness, but there is a notable absence of any systems with periods between \til 1
and 3 hr
 \cite{whi85a,whi85b}.  Recently, \scite{king97} have investigated the formation of
neutron star LMXBs.  They found that in order to produce the relatively
large fraction of soft X-ray transients in the $\simlt$ 1-2 day range, the
secondaries must have 1.3\msun $\simlt M_2\simlt$ 1.5\msun\ at the onset of mass
transfer and be significantly nuclear evolved (provided that the SN kick-velocity is small compared to the pre-SN orbital velocity).  This mass
range ensures that the mass transfer rates driven by angular momentum loss are below the critical rate needed for SXT behaviour in the required
fraction of neutron star LMXBs.  The large initial masses then account for the rarity of any short $P \simlt 3$
hr systems.  If
indeed the 2 hr modulation of GS1826-24 is confirmed as orbital in origin,
this would make it of great interest within the context of transient
neutron star LMXB formation and evolution. 

However, in many respects GS1826-24 does show similarities to other neutron star
binaries with comparable periods.  The X-ray bursters X1636-536
and X1735-444 exhibit optical modulations of a similar amplitude on their orbital
periods (3.80 and 4.65
hr respectively), plus irregular variability on
somewhat shorter timescales (\pcite{whi95} and references therein), although unlike GS1826-24 these systems are not transient on a timescale of decades.
The optical emission of LMXBs is dominated by the reprocessed X-rays from the accretion disc and companion.  The
origin of the underlying sinusoidal modulation is interpreted as the
varying contribution from the X-ray heated face of the companion star \cite{vP95a}. 
As for the irregular variability, this is probably caused by changes in the spatial distribution of the reprocessing material in the accretion
disc and/or fluctuations in the central X-ray luminosity, as suggested in the case of X1636-536 \cite{vP90b}.  Moreover, GS1826-24 has a high $L_X/L_{opt}(\sim500)$,
very similar to that of the compact 41 min
binary X1627-673, but lower than the $L_X/L_{opt}\sim700$ of the 50 min binary X1916-053, which is a higher inclination dipping source (once again these
are persistent sources).  Since this ratio
is related in part to the physical size of the system and the disc  reprocessing area available, the similarity supports the
hypothesis that GS1826-24 is also a relatively compact system.
 
 
Furthermore, the non-detection of GS1826-24 prior to 1988 is characteristic
of the observed variability of the transient bursting systems X0748-678,
X1658-298 and X2129+470 \cite{whi95}, which have either been detected over
many years and then gone into quiescence for a similar period of time or
vice-versa.  These transients all have known periods in the range
3.82-7.1hrs (see Table \ref{tab:sptb}).

Clearly, further high-speed optical photometric monitoring is required in order to confirm the stability of the 2hr modulation and hence
its orbital origin.   With only a short {\it ROSAT} lightcurve published to date \cite{barr95}, confirmation might be possible from a longer
X-ray observation. However, the low-amplitude of the observed modulation ($0.06^m$) implies a low-inclination system ($< 70 \degree$), and hence X-ray dipping behaviour is
unlikely to be seen.

\section{ACKNOWLEDGMENTS}
We are grateful to Erik Kuulkers for comments on an earlier version of this paper.  We thank Fran\c{c}ois van Wyck (SAAO), Chris Benn and Peter Sorensen (La Palma) for their support at the telescopes.  L.H. acknowledges the support
of a PPARC studentship.  The WHT is operated on the island of La Palma by the Royal Greenwich Observatory in the Spanish Observatorio del
Roque de Los Muchachos of the Institutio de Astrof\'{\i}sica de Canarias.
\bibliographystyle{mnras}

\end{document}